\documentclass[graybox]{svmult}

\usepackage{mathptmx}       % selects Times Roman as basic font
\usepackage{helvet}         % selects Helvetica as sans-serif font
\usepackage{courier}        % selects Courier as typewriter font
\usepackage{type1cm}        % activate if the above 3 fonts are
\usepackage{makeidx}         % allows index generation
\usepackage{graphicx}        % standard LaTeX graphics tool
\usepackage{multicol}        % used for the two-column index
\usepackage[bottom]{footmisc}% places footnotes at page bottom

\usepackage{latexsym,amssymb,rotating}

\makeindex             % used for the subject index
\begin{document}

\title*{Molecular modelling and simulation of electrolyte solutions, biomolecules, and wetting of component surfaces}
\titlerunning{Molecular simulation of electrolytes, biomolecules, and component surfaces (MOCOS)}
\author{M.\ Horsch,$^{1,\,\star}$ S.\ Becker,$^1$ J.\ M.\ Castillo,$^1$ S.\ Deublein,$^1$ A.\ Fr\"oscher,$^1$ S.\ Reiser,$^1$ S.\ Werth,$^1$ J.\ Vrabec,$^2$ and H.\ Hasse$^1$}
\authorrunning{Horsch, Becker, Castillo, Deublein, Fr\"oscher, Reiser, Werth, Vrabec, and Hasse}
\institute{$^1$TU Kaiserslautern, Lehrstuhl f\"ur Thermodynamik, Erwin-Schr\"odinger-Str.\ 44, 67663 Kaiserslautern \and $^2$Universit\"at Paderborn, Lehrstuhl f\"ur Thermodynamik und Energietechnik, Warburger Str.\ 100, 33098 Paderborn \and $^\star$Corresponding author: Martin Horsch, phone: +49 631 205 4028, \email{martin.horsch@mv.uni-kl.de}}
\maketitle

\abstract{
   Massively-parallel molecular dynamics simulation is applied to
   systems containing electrolytes, vapour-liquid interfaces, and
   biomolecules in contact with water-oil interfaces. Novel molecular
   models of alkali halide salts are presented and employed for the
   simulation of electrolytes in aqueous solution.
   The enzymatically catalysed hydroxylation of oleic acid is investigated
   by molecular dynamics simulation taking the internal degrees of
   freedom of the macromolecules into account.
   Thereby, Ewald summation methods are used to compute the
   long range electrostatic interactions.
   In systems with a phase boundary, the dispersive
   interaction, which is modelled by the Lennard-Jones potential
   here, has a more significant long range contribution than in
   homogeneous systems.
   This effect is accounted for by implementing the Jane\v{c}ek cutoff correction
   scheme. On this basis, the HPC infrastructure at the Steinbuch
   Centre for Computing was accessed and efficiently used, yielding
   new insights on the molecular systems under consideration.
}

\section{Introduction}
\label{sec:introduction}

Molecular simulation provides detailed information on processes on a
level which is other\-wise inaccessible, on the basis of physically
realistic models of the intermolecular interactions. However, despite
all efforts to keep these models as simple as possible, such
simulations are extremely time consuming and belong to the most
demanding applications of high performance computing.

The scientific computing project \textit{``Molecular simulation of
static and dynamic properties of electrolyte systems, large molecules
and wetting of component surfaces''} (MOCOS) aims at advancing the state
of the art by developing the molecular simul\-ation codes \textit{ms2}
\cite{DESLGGMBHV11} as well as \textit{ls1 mardyn} \cite{BBV11}, and
by applying advanced simulation algorithms to complex molecular
systems, employing the HPC resources at the Steinbuch Centre for
Computing (SCC) in Karlsruhe. The following peer-reviewed
publications, which have appeared (or will appear) in internationally
accessible journals, contribute to the MOCOS project in various
respects:
\begin{itemize}
   \item{} S.\ Deublein, S.\ Reiser, J.\ Vrabec, H.\ Hasse, \textit{A set of molecular models for alkaline-earth cations in aqueous solution}, J.\ Phys.\ Chem.\ B \textbf{116}(18), 5448 (2012)
   \item{} S.\ Deublein, S.\ Reiser, J.\ Vrabec, H.\ Hasse, \textit{Solute and solvent dynamics in aqueous solutions of alkali halide salts: A molecular simulation study using a con\-sistent force field}, J.\ Phys.\ Chem., submitted (2013)
   \item{} Y.-L.\ Huang, T.\ Merker, M.\ Heilig, H.\ Hasse, J.\ Vrabec, \textit{Molecular modeling and simulation of vapor-liquid equilibria of ethylene oxide, ethylene glycol, and water as well as their binary mixtures}, Ind.\ Eng.\ Chem.\ Res.\ \textbf{51}(21), 7428 (2012)
   \item{} S.\ Pa\v{r}ez, G.\ Guevara Carri\'on, H.\ Hasse, J.\ Vrabec, \textit{Mutual diffusion in the ternary mixture water + methanol + ethanol and its binary subsystems}, Phys.\ Chem.\ Chem.\ Phys.\ \textbf{15}(11), 3985 (2013)
   \item{} S.\ Reiser, N.\ McCann, M.\ Horsch, H.\ Hasse (2012), \textit{Hydrogen bonding of ethanol in supercritical mixtures with CO$_\mathit{2}$ by $^\mathit{1}$H NMR spectroscopy and molecular simulation}, J.\ Supercrit.\ Fluids \textbf{68}, 94 (2012)
   \item{} J.\ Walter, J.\ Sehrt, J.\ Vrabec, H.\ Hasse, \textit{Molecular dynamics and experimental study of conformation change of poly(N-isopropylacrylamide)-hydrogels in mixtures of water and methanol}, J.\ Phys.\ Chem.\ B \textbf{116}(17), 5251 (2012)
   \item{} S.\ Werth, S.V.\ Lishchuk, M.\ Horsch, H.\ Hasse, \textit{The influence of the liquid slab thickness on the planar vapor-liquid interfacial tension}, Phys.\ A \textbf{392}(10), 2359 (2013)
\end{itemize}
The present article addresses the three main topics of the MOCOS
project by presenting recent scientific results which were
facilitated by the computing resources at SCC: Section
\ref{sec:electrolytes} discusses molecular model development for ions
in aqueous solution and Section \ref{sec:cytochrome} reports on
molecular simulation of large molecules, i.e.\ of enzymes and fatty
acides. Interfacial phenomena such as the wetting behaviour of
fluids in contact with a solid surface are discussed in Section
\ref{sec:interfaces} and a conclusion is given in Section
\ref{sec:conclusion}.

\section{Molecular modelling of aqueous electrolyte solutions}
\label{sec:electrolytes}

\subsection{Force field development for alkali cations and halide anions}

Aqueous electrolyte solutions play an important role in many industrial applications and natural processes. The general investigation of electrolyte solutions is, hence, of prime interest. Their simulation on the molecular level is computationally expensive as the electrostatic long range interactions in the solution have to be taken into account by time consuming algorithms, such as Ewald summation \cite{Ewald21}.

The development of ion force fields in aqueous solutions is a challenging task because of the strong electrostatic interactions between the ions and the surrounding water molecules. Previously published parameterization strategies for the adjustment of the ion force fields of all alkali cations and halide anions yield multiple parameter sets for a single ion \cite{HMN09, RH11} or are based on additional assumptions, e.g.\ the consideration of the aqueous alkali halide solution as a binary mixture of water and cations as well as anions, respectively \cite{GCJBW11}.

The recent study of Deublein \textit{et al.}\ \cite{DVH12}, however, succeeded in obtaining one unique force field set for all alkali and halide ions in aqueous solution. Thereby, the ions are modelled as Lennard-Jones (LJ) spheres with a point charge ($\pm 1$ $e$) in their centre of mass. Hence, the ion force fields have two adjustable parameters, namely the LJ size parameter $\sigma$ and the LJ energy parameter $\epsilon$. The $\sigma$ parameter of the ions was adjusted to the reduced liquid solution density. The LJ energy parameter showed only a minor influence on the reduced density and was estimated to be $\epsilon = 100$ K for all anions and cations \cite{DVH12}.

In the present work, the influence of the LJ energy parameter on the self-diffusion coefficient of the alkali cations and the halide anions in aqueous solutions as well as the position of the first maximum of the radial distribution function (RDF) of water around the ions was investigated systematically. Based on these results, a modified value is proposed for the LJ energy parameter.

The new $\epsilon_{i}$ parameter of the ion force fields is determined by a two step para\-metrization strategy. First, the LJ energy parameter of the ion force fields is adjusted to the self-diffusion coefficient of the ions in aqueous solution. Subsequently, the dependence of the position of the first maximum in the RDF of water around the ions on $\epsilon_{i}$ is used to restrict the parameter range derived by considering the self-diffusion coefficient.
% The $\epsilon_{i}$ value within this parameter range which possesses the smallest deviation of the simulation results for $r_{\mathrm{max,1}}$ from the experimental data is determined as new LJ energy parameter of the ion force fields.

% The present study of the self-diffusion coefficient $D_i$ and the first maximum $r_{\mathrm{max,1}}$ of the RDF $g_{i-\mathrm{O}}(r)$ of water around the ions indicates that $\epsilon_{i} / k_{\mathrm{B}} = 200$~K is a reasonable choice for all alkali cations as well as all halide anions.

\subsection{Methods and simulation details}

In the present study, the self-diffusion coefficient of the ions in aqueous solutions is determined in equilibrium molecular dynamics by the Green-Kubo formalism. In this formalism, the self-diffusion coefficient is related to the time integral of the velocity autocorrelation function \cite{Gubbins72}. The radial distribution function $g_{i-\mathrm{O}}(r)$ of water around the ion $i$ is defined by
\begin{equation}
g_{i-\mathrm{O}}(r)=\frac{\rho_{\mathrm{O}}(r)}{\rho_{\mathrm{O},\mathrm{bulk}}}
\end{equation}
where $\rho_{\mathrm{O}}(r)$ is the local density of water as a function of the distance $r$ from the ion $i$ and $\rho_{\mathrm{O},\mathrm{bulk}}$ is the density of water molecules in the bulk phase. In this case, the position of the water molecules is represented by the position of the oxygen atom. The radial distribution functions are evaluated by molecular dynamics (MD) simulation as well.

The calculation of the self-diffusion coefficient by the Green-Kubo formalism and the evaluation of the RDF is time and memory consuming. The determination of $D_i$ and $g_{i-\mathrm{O}}(r)$ in electrolyte solutions is considerably more expensive due to additional time consuming algorithms, e.g.\ Ewald summation \cite{Ewald21}, required for permitting a truncation of the long range electrostatic interactions; however, it should be noted that a completely explicit evaluation of all pairwise interactions would be even much more expensive. 
% The proceeding for the simulation of the self-diffusion coefficient and the radial distribution function is identical. However, they are not calculated simultaneously as both algorithms contribute to the total computing time. For the evaluation of the RDF, only small systems (N = 1000) are needed. The simulation of larger systems leads to significant higher computing times without further achievements. In contrast, the Green-Kubo formalism requires large numbers of ions in the solutions for the determination of $D_i$ with acceptable uncertainties. As, at the same time, infinite dilution is aspired, the systems have to be relatively large (N = 4500).

In a first step, the density of the aqueous alkali halide solution was determined in an isobaric-isothermal (\textit{NpT}) MD simulation at the desired temperature and pressure. The resulting density was used in a canonical (\textit{NVT}) MD simulation at the same temperature, pressure and composition of the different alkali halide solutions. In this run, 
the self-diffusion coefficient of the ions as well as the radial distribution function, respectively, was determined. In case of the calculation of $D_i$, the MD unit cell with periodic boundary conditions contained $N$ = 4 500 molecules, both for the \textit{NpT} and the \textit{NVT} simulation run.
% The simulation volume included 4420~water molecules, 40~alkali cations and 40~halide anions. The sampling length of the velocity auto-correlation function was set to 11~ps and the time span between the origins of two auto-correlation functions was chosen such that all auto-correlation functions have decayed at least to $1/e$ of their normalized value.

For the evaluation of the RDF, there were $N$ = 1 000 molecules in the simulation volume, i.e.\ 980 water molecules, 10 alkali cations and 10 halide anions. The radial distribution function was sampled in the \textit{NVT} simulation within a cutoff radius of 15~\r{A} with 500 bins.

\subsection{Self-diffusion coefficients and radial distribution functions}

The self-diffusion coefficient $D_i$ and the position of the first maximum $r_{\mathrm{max,1}}$ of the RDF $g_{i-\mathrm{O}}(r)$ of water around the ions was investigated for all alkali cations and halide anions in aqueous solution for $\epsilon = 200$ K. These data were calculated at high dilution so that correlated motions and ion pairing between the cations and anions were avoided. Hence, $D_i$ and $r_{\mathrm{max,1}}$ are independent on the counter-ion in solution.

\begin{figure}[t]
%\sidecaption
\center
\includegraphics[width=5.9cm]{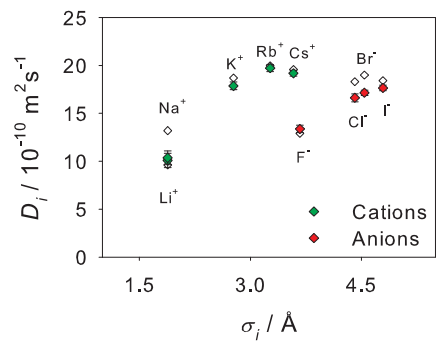}
\caption{Self-diffusion coefficient of alkali cations and halide anions in aqueous solutions ($x_{\mathrm{S}}=0.018$~mol/mol) at $T=298.15$~K and $p=1$~bar. Simulation results (full symbols) are compared to experimental data~\cite{ML89} (empty symbols).}
\label{Adjust_plot_SD_cation_anion}   
\end{figure}

The results for the self-diffusion coefficient $D_i$ are shown in Fig.\ \ref{Adjust_plot_SD_cation_anion}. The overall agreement of the simulation results with the experimental data is excellent. The deviations are below 10~\% for all ions, except for the sodium cation, where the deviation is about 20~\%.

These simulation results also follow the qualitative trends from experiment, i.e.\ $D_i$ increases with cation and anion size, respectively. This ion size dependence is directly linked to the electrostatic interaction between the ions and water. In aqueous solution, the cations and anions are surrounded by a shell of electrostatically bonded water molecules (hydration shell). The ions diffuse together with their hydration shell within the bulk water. For small ions, the hydration shell is firmly attached to the ion. Hence, the effective radius, that typically dominates ion motion, is larger for smaller ions than for larger ions, where the hydration shell is less pronounced.

\begin{figure}[h]
%\sidecaption
\center
\includegraphics[width=5.9cm]{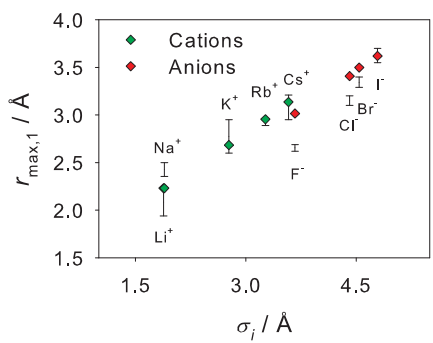}
\caption{Position of the first maximum $r_{\mathrm{max},1}$ of the RDF $g_{i-\mathrm{O}}(r)$ of water around the alkali cations and halide anions in aqueous solutions ($x_{\mathrm{S}}=0.01$~mol/mol) at $T=293.15$~K and $p=1$~bar. Present simulation data (full symbols) are compared to the range of experimental $r_{\mathrm{max,1}}$ data~\cite{OR93, FPWNSM96} (vertical lines).}
\label{Adjust_plot_Rmax_cation_anion}      
\end{figure}

The results of $r_{\mathrm{max,1}}$ are shown in Fig.\ \ref{Adjust_plot_Rmax_cation_anion}. In case of the alkali cations, the simulation results are within the range of the experimental data, except for Na$^+$ where the deviation from the experimental data is 5.3\%.
For the halide anions, only the simulation result for $r_{\mathrm{max,1}}$ of the RDF of water around the iodide anion is within the range of the experimental data. The deviations of the simulation results for $r_{\mathrm{max,1}}$ from the experimental data are 12.1 \% around F$^-$, 6.5 \% around Cl$^-$, and 2.9 \% around Br$^-$.

% The position of the first maximum in the RDF of water around the ions increases with increasing size both of the alkali cations and halide anions. The dependence of the simulation results for $r_{\mathrm{max,1}}$ on the size of the ions is nearly identical for the cations and the anions, cf.~\ref{Adjust_plot_Rmax_cation_anion}.

Comparing $D_i$ and $r_{\mathrm{max,1}}$ of the cesium cation and the fluoride anion, which have almost the same size, it can be seen that Cs$^+$ diffuses faster in aqueous solution and the water molecules of the hydration shell around F$^-$ are closer to the ion. This can be attributed to the different orientations of the water molecules around the oppositely charged ions.
% In case of the positive cesium ion, the negatively charged oxygen atom of the water molecule points towards the ion and hence the two positive hydrogen atoms face into the opposite direction. As the equally charged hydrogen atoms of neighboring water molecules reject each other, the water molecules are not able to form a strongly attached hydration shell around the cesium cation.
The water molecules are able to build a stronger attached hydration shell around the fluoride ion which is closer to the ion.

% In comparison, predominantly only one of the two hydrogen atoms of the water molecule points towards the negatively charged fluoride anion~\cite{Impey1983}. The second hydrogen atom, which turns away from the ion, is attracted by the oxygen atom of a neighboring water molecule. Hence, the water molecules are able to build a stronger attached hydration shell around the fluoride ion which is closer to the ion.

\subsection{Computational demands}

The molecular simulations in Section \ref{sec:electrolytes} were carried out with the MPI based molecular simulation program \textit{ms2}, which was developed in our group, cf.\ Deublein \textit{et al.}\ \cite{DESLGGMBHV11}. The total computing time for determining the self-diffusion coefficient of ions in aqueous solutions was 138 hours on 36 CPUs (48 hours for the \textit{NpT} run and 90 hours for the \textit{NVT} run). For these simulations a maximum virtual memory of 1.76~GB was used.

For the evaluation of the radial distribution function of water around the ions a total computing time of 31 hours on 32 CPUs (10 hours for the \textit{NpT} run and 21 hours for the \textit{NVT} run) was required.

\section{Simulation of biomolecules with internal degrees of freedom}
\label{sec:cytochrome}

\subsection{Catalysed hydroxylation of unsaturated fatty acids}

Producing polymers from regenerative feedstock is a highly interesting alternative to polymers made of naphtha. The first step for obtaining biopolymers is the synthesis and study of all possible basic materials that can be used as building blocks. Such materials can for instance be obtained by an enzymatically catalysed reaction of unsaturated fatty acids to dihydroxy-fatty acids. 

In nature, there are only mixtures saturated and unsaturated fatty acids rather than the pure compounds. Therefore, separation -- before or after the reaction -- is necessary to obtain pure products, e.g.\ by chromatography, using hydrotalcite as adsorbent and a mixture of water and isopropanol as solvent.

Cytochrome P450 monooxygenase, an enzyme well-known for catalysing the hydroxylation of organic molecules, is a suitable catalyst for this process as well. The critical aspect of the catalytic reaction is the contact between the heme group, which contains the active centre of the enzyme, and the double bonds in the carbon chain of the fatty acid. The enzyme is denaturalized in the presence of organic molecules, and is only active in an aqueous phase. Fatty acids have a small dipole moment and are almost insoluble in water. So how does the contact between the active centre of the enzyme and the fatty acid take place?

A series of molecular simulations was conducted to learn more about the distribution of molecules around the enzyme and the behaviour of the system at different conditions. For this purpose, the mixing behaviour of the systems fatty acid + water, fatty acid + water + isopropanol, and fatty acid + water + cytochrome P450 was investigated. In particular, it is relevant to know whether the fatty acid builds micelles in the water-isopropanol solvent, or how the enzyme catalyses the reaction despite the different phase behaviour of the solvent and the fatty acid. The fatty acid of interest in the present work is oleic acid.

\subsection{Simulation details}

Molecular simulation of biological systems poses a challenge to scientific computing. The most important limitation in molecular simulation is the system size. As the number of molecules in the simulation box $N$ increases, the computing time required for the simulation increases with $O(N^2)$ if it is implemented in a naive way. The reason for this steep dependence is found in the computation of pair potentials, so that the distance and energy between interacting atoms needs to be computed at every simulation step.

One option for decreasing the simulation time consists in following a coarse-graining approach \cite{MuellerPlathe2002}. By coarse-graining, a group of atoms is modelled as a single interaction centre, reducing the number of pair interactions that have to be calculated. As we are interested in obtaining detailed atomistic information, we prefer to use a full atomistic model. The model we selected is OPLS \cite{JMT96}, which has been successfully used to simulate a large variety of biological systems. The model for the heme group, inexistent in the original OPLS force field, was taken from a recent parameterization of this group compatible with the OPLS force field \cite{GSB06}. In these models, repulsion-dispersion interactions are treated with LJ potentials, while electrostatic interactions are taken into account considering point partial charges at the atomic positions. Internal molecular degrees of freedom are modelled via bond, bend, and torsion potentials.

% Another inconvenient of large molecular simulations, as typically found in biological systems, is that sophisticated simulation techniques are required for efficiency. For example, the most accepted method to calculate long range interactions, such as electrostatic interactions, is the Ewald summation technique \cite{Ewald21}. Unfortunately, this technique becomes inefficient when the system contains more than 10 000 interaction centres \cite{FS02}, so other methods such as Particle Mesh Ewald \cite{TM00} need to be used instead.

\begin{figure}[b]
%\sidecaption
\center
\includegraphics[width=6.85cm]{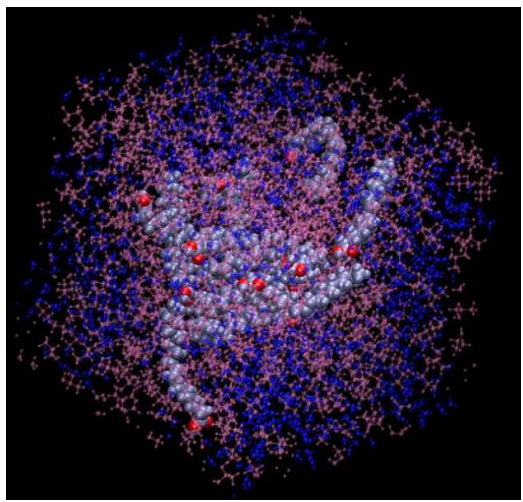}
\caption{Cluster of oleic acid molecules in water-isopropanol solution. Dark blue, water; violet, isopropanol; light blue, oleic acid. The acid group of the oil molecules is painted in red.}
\label{fig:mocosfig3}      
\end{figure}

The cytochrome P450 enzyme contains more than 7 000 atoms, and together with the fatty acid and the solvent molecules, we need to simulate up to 180 000 atoms. Molecular simulation of such an enormous system would need months of simulation time unless we use advanced parallel simulation programs. For the present simulations of molecular biosystems, the \textit{GROMACS 4.6} MD program was used \cite{PSSLBASSKSHL13}. At the beginning of the simulation, the molecules were arranged in a randomized fashion within a periodic simulation box, and the energy was minimized. This procedure for generating a starting configuration attempts to mimic experimental conditions where initially the solutions are vigorously mixed. Then every molecule is assigned a random velocity (corresponding to the temperature of the system), and a short simulation of 2 ps is carried out for an initial equilibration. Subsequently, a long simulation (over a minimum simulation time of 20 ns) is run, where the properties of interest are calculated.

The molecular positions are updated using a leapfrog algorithm to integrate Newton's equations of motion. We truncate the LJ potential at a cut-off radius of $r_\mathrm{c} = 1.5$ nm, and use the particle-mesh Ewald summation method \cite{TM00} to calculate electrostatic interactions. Temperature is controlled by velocity rescaling with a stochastic term, and pressure with a Berendsen barostat. More details about these simulation techniques can be found elsewhere \cite{FS02}.

Our simulations are performed under conditions for which reliable experimental data are available, i.e.\ at a temperature of 298 K and a pressure of 100 kPa. For the water-oleic acid system, we use a fatty acid mass fraction of 60 \% and two different simulation boxes with approximately the same volume: A cubic box with $V$ = $(7.14 nm)^3$, and a ortho\-rhombic box of the dimension 5$\times$15$\times$5 nm$^3$. For the isopropanol-water-oleic acid system, we added 10 mg/ml of oleic acid to a mixture water/isopropanol with a mass fraction of 60 \% (for isopropanol) in a cubic simulation box of (12 nm)$^3$. Finally, the oil/water concentration in the cytochrome P450 + water + oil system corres\-ponded to a volume fraction of 30 \% (for oleic acid) in a cubic box of (10 nm)$^3$, wherein a single cytochrome P450 enzyme was placed.

\begin{figure}[h]
%\sidecaption
\center
\includegraphics[width=6.85cm]{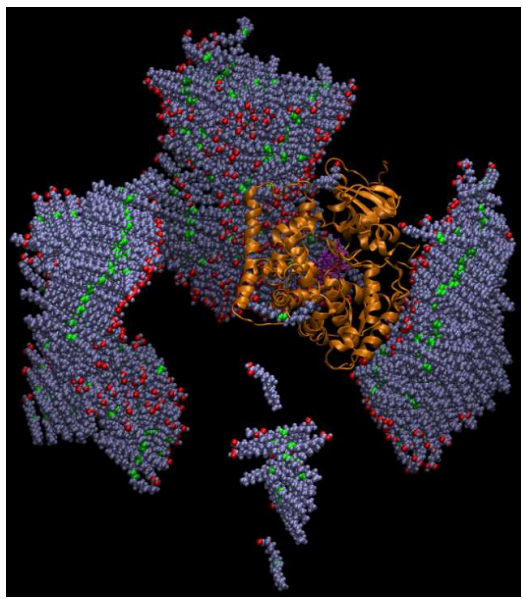}
\caption{Cytochrome P450 in water-oleic acid solution. Cytochrome P450 is painted in orange; the heme group, in violet; the oil molecules, in light blue. The acid group of the oil molecules is painted in red, the double bond in green. For clarity, water has been removed in the picture.}
\label{fig:mocosfig4}      
\end{figure}

\subsection{Simulation results}

Molecular simulations of oil in the presence of water show a swift phase separation, as it is expected from experimental observation. The phase separation takes place only after a few picoseconds. Independently of the shape of the simulation box, the oleic acid forms a bilayer, while other possible structures are not observed. Oil molecules form a ordered phase, where their acid groups point in the direction of the water phase, and their hydrocarbon tails are aligned. This is a consequence of the well known fact that the acid group is hydrophilic while the hydrocarbon tail is hydrophobic. 

When we add isopropanol to the water-oil mixture, the situation changes, as shown in Fig.\ \ref{fig:mocosfig3}. Water and isopropanol are miscible, and their behaviour is similar as in a pure water-isopropanol solution.
% Oil can either form small, disordered clumps of molecules, or have single molecules in solution.
Oil molecules are mostly surrounded by isopropanol, which can be easily inferred from the radial distribution functions. On the other hand, the double bonds in the oil clusters tend to be in contact with each other.

Oil in the presence of water and cytochrome P450 behaves similarly as in water-oil solutions, cf.\ Fig.\ \ref{fig:mocosfig4}. After a few nanoseconds a clear phase separation takes place, where oil forms an ordered, separate phase. Cytochrome P450 stays solvated in water and in contact with the oil phase only at specific points. At the beginning of the simulation there are several oil molecules in the vicinity of cytochrome P450 at favourable contact sites. These molecules maintain their contact with the enzyme during the whole simulation in a position where an interaction with the active centre (which is situated within the heme group) is possible.

\subsection{Computational demands}

The computational demands of our simulations were highly dependent on the type of simulation. For the energy minimization simulations, it was sufficient to use a single processor for several minutes. For equilibration, we used 16 CPUs running in parallel for a maximum of 3.5 hours. The most demanding simulations during production required 256 CPUs running for 180 hours and a virtual RAM of 2.8 GB.

\section{Vapour-liquid interfaces and wetting of component surfaces}
\label{sec:interfaces}

\subsection{Planar vapour-liquid interfaces}

For simulating vapour-liquid coexistence on the molecular level, a long range correction is needed to obtain accurate results. Thereby, the dispersive contribution to the potential energy $U_i$ of a molecule $i$
\begin{equation}
 U_i = \sum_{r_{ij}<r_c} u_{ij} + U_{i}^{LRC}
\end{equation}
is calculated as a sum of the explicitly computed part and the long range correction (LRC), where the latter consists of a summation over $N_s$ slabs, employing a periodic boundary condition. The LRC is only applied in the direction normal to the planar interface, corresponding to the $y$ direction here
\begin{equation}
 U_{i}^{LRC} = \sum_{k}^{N_s} \Delta u_{i,k}^{LRC} (y_i,y_k),
\end{equation}
given that the system is homogeneous in the other directions.
The correction terms $\Delta u_{i,k}^{LRC} (y_i,y_k)$ for the slabs are calculated as an integral over the slab volume. The resulting term for the potential energy only depends on the distance between the slabs $r$, the density $\rho$ and the thickness $\Delta y$ of the slabs \cite{Janecek06}.
% \begin{equation}
%   \Delta u_{i,k}^{LRC} (y_i,y_k) = \Delta y \rho (y_k) 4 \pi \epsilon \left[ \frac{1}{5} \left( \frac{\sigma}{r} \right)^{10} - \frac{1}{2} \left( \frac{\sigma}{r} \right)^{4}\right].
% \end{equation}
% If the distance between the two slabs $i$ and $k$ is smaller the the cutoff radius, the cutoff radius $r_\mathrm{c}$ is used instead of $r$.
The slabwise interaction is computed in pairs, so that $\frac{1}{2} N_s^2$ individual contributions have to be computed.
% The parallelization is implemented by distributing the slab interaction over the different threads, i.e.\ according to the pattern from Fig.\ \ref{fig:werth1}.

% \begin{figure}[hbt]
% \centering
% %\sidecaption
% % Use the relevant command for your figure-insertion program
% % to insert the figure file.
% % For example, with the graphicx style use
% \includegraphics[width=7cm]{mocosfig5}
% %
% % If no graphics program available, insert a blank space i.e. use
% %\picplace{5cm}{2cm} % Give the correct figure height and width in cm
% %
% \caption{Distribution of the slab interaction to the different threads. Each color represents a thread. The distribution aims to a homogeneous load for every thread.}
% \label{fig:werth1}       % Give a unique label
% \end{figure}

\begin{figure}[t]
\centering
%\sidecaption
% For example, with the graphicx style use
\includegraphics[width=5.9cm]{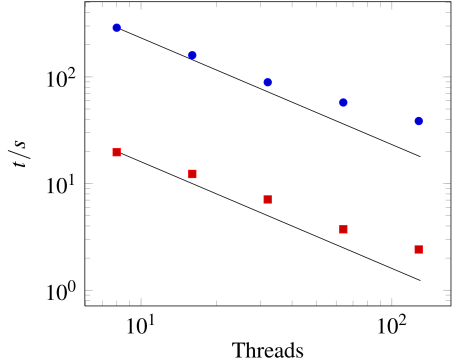}
%
% If no graphics program available, insert a blank space i.e. use
%\picplace{5cm}{2cm} % Give the correct figure height and width in cm
%
\caption{Strong scaling with $N$ = 256 000 2CLJ particles and 512 slabs. The blue dots denote the run time of the whole program, while the red dots only denote the run time of the long range correction. The solid lines represent ideally linear scaling.}
\label{fig:werth2}       % Give a unique label
\end{figure}

Usually, the number of threads is much smaller than the number of slabs. For a scaling test, a system with $N$ = 256 000 two-centre Lennard-Jones (2CLJ) particles was simulated with different numbers of threads and 512 slabs. Fig.\ \ref{fig:werth2} shows the results for the strong scaling behaviour.

For small numbers of threads, the program scales almost ideally, i.e.\ linearly. For larger numbers of threads, the communication between the threads and the decomposition of the particles becomes more time consuming. Eventually, the long range correction requires even more communication between the threads than the rest of the program, since the density profile has to be sent to every thread before the slab interaction begins. After the slab interaction has been calculated, the correction terms for the potential energy, the force and the virial have to be distributed to every thread.

On the basis of the long range correction discussed above, vapour-liquid equi\-libria were considered by explicit MD simulation of the fluid phase coexistence (i.e.\ including an interface), so that the vapour-liquid surface tension $\gamma$ could be computed following the virial route. For the (single centre) LJ fluid, a high precision was obtained by simulating $N$ = 300 000 particles over roughly a million time steps, resulting in very small statistical uncertainties. From the resulting surface tension values, the regression term
\begin{equation}
 \gamma = 2.94 \left( 1- \frac{T}{T_c}\right)^{1.23} \frac{\epsilon}{\sigma^2}
 \label{eq:regression}
\end{equation}
was obtained, where the LJ critical temperature is given by P\'erez Pellitero \textit{et al.}\ \cite{PUOM06} as $T_c$ = 1.3126 $\epsilon$. 

% \begin{figure}[hbt]
% \centering
% %\sidecaption
% % Use the relevant command for your figure-insertion program
% % to insert the figure file.
% % For example, with the graphicx style use
% \includegraphics[width=7cm]{mocosfig7.eps}
% %
% % If no graphics program available, insert a blank space i.e. use
% %\picplace{5cm}{2cm} % Give the correct figure height and width in cm
% %
% \caption{Reduced surface tension over the reduced temperature. Comparison of the results of different authors. The solid line represents the regression.}
% \label{fig:werth3}       % Give a unique label
% \end{figure}

\subsection{Curved vapour-liquid interfaces}

\begin{figure}[h]
		\begin{center}
			\includegraphics[width=8.75cm]{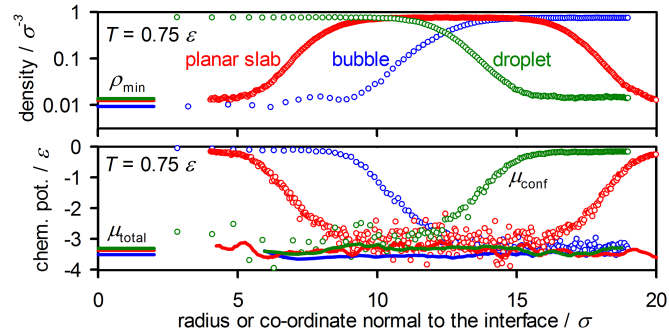}
			\caption{Density profiles (top) and chemical potential profiles (bottom) for nanoscopic vapour-liquid interfaces of the TSLJ fluid from canonical ensemble MD simulation of systems containing planar liquid and vapour slabs, a liquid drop (surrounded by a supersaturated vapour), or a gas bubble (surrounded by a subsaturated liquid), respectively. The chemical potential profile includes information on both the configurational contribution (circles) and the total chemical potential (solid lines), as a function of the characteristic spatial coordinate of the system. The solid lines in the bottom left part of the two diagrams indicate the density of the vapour phase (top) as well as the average value of the total chemical potential (bottom).}
		\label{fig:bedhotiya}
		\end{center}
\end{figure}

The influence of curvature on vapour-liquid interfacial properties was examined for the trunacted-shifted Lennard-Jones fluid (TSLJ), with a cutoff radius of $r_\mathrm{c}$ = 2.5 $\sigma$, by simulating both curved (i.e.\ bubble and droplet) as well as planar interfaces by MD simulation of the canonical ensemble at a reduced temperature of $T = 0.75$ $\epsilon$. For the system with planar symmetry, half of the simulation volume was filled with vapour and the other half with liquid, using a simulation box with a total elongation of 17 $\sigma$. The density profile from this simulation was compared with that of a bubble as well as a droplet with an equimolar radius of $R_\rho$ = 12.5 $\sigma$, cf.\ Fig.\ \ref{fig:bedhotiya}.
A system containing a bubble is subsaturated, whereas the fluid phases coexisting at the curved interface of a droplet are supersaturated with respect to the thermodynamic equilibrium condition for the bulk phases coexisting at saturation. In agreement with this qualitative statement from phenomenological thermodynamics, the vapour density (i.e.\ the minimal value from the density profile) was found to be smaller for the system containing a bubble ($\rho''$ = 0.0097 $\sigma^{-3}$) and larger for the system containing a droplet (0.0140 $\sigma^{-3}$), whereas the vapour density over the planar interface was 0.0127 $\sigma^{-3}$. 

Analogous results were obtained for the chemical potential $\mu^\mathrm{total} = \mu^\mathrm{id} + \mu^\mathrm{conf}$, which was determined as the sum of the ideal contribution $\mu^\mathrm{id} = T \ln \rho$ and the configurational contribution $\mu^\mathrm{conf}$. The Widom test particle method \cite{Widom82} was implemented to compute a profile of $\mu^\mathrm{conf}$, which in equilibrium is complementary to the logarithmic density profile, yielding an approximately constant profile for the total chemical potential, cf.\ Fig.\ \ref{fig:bedhotiya}. For the droplet, an average value of $\mu^\mathrm{total}$ = $(-3.31 \pm 0.04)$ $\epsilon$ was determined, in comparison to $(-3.51 \pm 0.02)$ $\epsilon$ for the bubble and $(-3.37 \pm 0.02)$ $\epsilon$ for the planar slab. This shows that in the present case, curvature effects are more significant for a gas bubble than for a liquid droplet of the same size, suggesting that the surface tension of the bubble is larger than that of the droplet.

\subsection{Temperature dependence of contact angles}

The research on the wetting behaviour of surfaces is an active field in materials science.
The goal is to reliably predict the wetting properties of component surfaces for design of components with new functional features. 
In this regard molecular simulations are particularly useful due to their high resolution. Moreover, molecular simulation permits the systematic investigation of the influence of particular parameters on the wetting behaviour. 

\begin{figure}[b]
		\begin{center}
			\includegraphics[width=6.85cm]{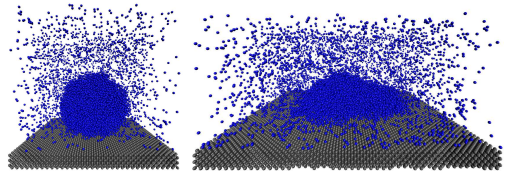}
			\caption{Snapshots of simulation runs at different fluid-solid interaction parameters $\zeta$ of 0.35 (left) and 0.65 (right) at a temperature
			 of 0.8 $\epsilon/k$.}
		\label{fig:SnapshotSD}
		\end{center}
\end{figure}

For the present study, the TSLJ potential was employed and the contact angle dependence on temperature for a non-wetting ($\theta>90^\circ$) and a partially wetting ($\theta < 90^\circ$) scenario, cf.\ Fig.\ \ref{fig:SnapshotSD} was investigated by MD simulation.
The solid wall was represented by a face centered cubic lattice with a lattice constant of $a = 1.55$ $\sigma$ and the (001) surface exposed to the fluid.
For the unlike interaction between the solid and the fluid phase the size and energy parameters $\sigma_\mathrm{sf} = \sigma_\mathrm{ff}$ and 
\begin{equation}
\epsilon_\mathrm{sf} = \zeta \epsilon_\mathrm{ff},
\label{bertehlot}
\end{equation}
respectively, were applied.

%######################################################################################################################################################

The simulations were carried out with the massively parallel MD program \textit{ls1 mardyn} \cite{BBV11}.
A total number of $N$ = 15 000 fluid particles was simulated in the \textit{NVT} ensemble.
In most cases, 64 processes were employed for parallel computation.
The efficient simulation of heterogeneous systems,
containing vapour-liquid interfaces and even a solid component, was
facilitated here by a dynamic load balancing scheme based on
$k$-dimensional trees \cite{BBV11}.
At the beginning of the simulation, the fluid particles were arranged on regular lattice sites in form of a cuboid.
The equilibration time of the three phase system was 2 ns, whereas the total simulation time was chosen to be 6 ns.
Periodic boundary conditions were applied in all directions, leaving a channel with a distance of at least 35 $\sigma$ between the wall and its periodic image, avoiding ``confinement effects" at near-critical temperatures \cite{OBG06}.
The contact angles were determined by the evaluation of density profiles averaged over a time span of 4 ns.
% , cf.\ Fig.\ \ref{fig:DichteProf}.

% \begin{figure}[hb]
		% \begin{center}
% 			\includegraphics[width=7cm]{mocosfig10}
			% \caption{Density profile of a sessile drop in direction radial from its center and normal to the wall, $x$ and $y$ in units of $\sigma$,
% 			 respectively. The colour depicts the density in units of $\sigma^{-3}$, averaged over 4 ns. The white circle is fitted to the equimolar radius.
			 % The black rectangle represents the wall. The effect of adsorption can clearly be seen on the right side of the drop.}
% 		\label{fig:DichteProf}
		% \end{center}
% \end{figure}

\begin{figure}[h]
		\begin{center}
			\includegraphics[width=6.85cm]{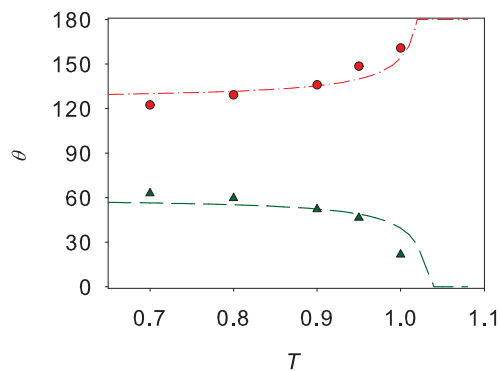}
			\caption{Simulation results and correlation for the dependence of the contact angle on the temperature for fluid-solid dispersive interaction strengths $\zeta$ of 0.35 (circles, dashed line) and 0.65 (triangles, dash-dotted line).}
		\label{CAonTemp}
		\end{center}
	\end{figure}

Sessile drops were studied at two different interaction parameters, $\zeta$ = 0.35 and 0.65, corresponding to partially wetting and non-wetting conditions. 
%The effect of the temperature on the contact angle was examined. 
The simulations were carried out in the temperature range between $0.7$ and $1$ $\epsilon$, covering nearly the entire regime of stable vapour-liquid coexistence for the TSLJ fluid, as indicated by the triple point temperature $T_3 \approx 0.65$ $\epsilon$, cf.\ van Meel \textit{et al.}\ \cite{MPSF08}, and the critical temperature $T_\mathrm{c} \approx 1.078$ $\epsilon$, cf.\ Vrabec \textit{et al.}\ \cite{VKFH06}.
It is well known that the respective wetting behaviour of the sytem, i.e.\ wetting or dewetting, is reinforced at higher temperatures, cf.\ Fig.\ \ref{CAonTemp}.
Close to the critical point of the fluid, the phenomenon of critical point wetting occurs \cite{Cahn77}, i.e.\ either the liquid or the vapour phase perfectly wets the solid surface
\begin{equation}
		\left| \cos\theta(T \rightarrow T_c) \right| \rightarrow 1.
		\label{cpWetting}
\end{equation}

\section{Conclusion}
\label{sec:conclusion}

Within the MOCOS project, a substantial progress was made regarding
software development for massively-parallel molecular simulation. In
particular, suitable algorithms for the long range contributions to
the intermolecular interaction were implemented in \textit{ms2}
(Ewald summation) and \textit{ls1 mardyn} (Jane\v{c}ek correction for
long range dispersive effects). In \textit{ls1 mardyn}, a
$k$-dimensional tree-based domain decomposition with dynamic load
balancing was implemented to respond to an uneven distribution of the
interaction sites over the simulation volume.
Furthermore, \textit{GROMACS} \cite{PSSLBASSKSHL13} was used to
simulate large molecules, taking their internal degrees of freedom
into account. In this way, systems with up to 300 000 interaction
sites were simulated for over 20 nanoseconds.

New models were developed for alkali cations and halide
anions in order to treat the dispersive interaction accurately in
molecular simulations of electrolye solutions. These models were
optimized by adjusting the LJ size parameter to the
reduced density of aqueous solutions and the energy parameter to
diffusion coefficients as well as pair correlation functions. On this
basis, quantitatively reliable results were obtained regarding the
structure of the hydration shells formed by the water molecules
surrounding these ions. Important qualitative structural information
on biomolecules at oil-water interfaces was deduced by analysing MD
simulations of the cytochrome P450 enzyme, and the critical wetting
behaviour as well as the properties of planar and curved vapour-liquid
interfaces were characterized by massively-parallel MD simulation
with the \textit{ls1 mardyn} program.

The scenarios disussed above show how
molecular modelling can today be applied to practically relevant
systems even if they exhibit highly complex structures and
intermolecular interactions, including significant long range
contributions. In the immediate future, molecular simulation as an
application of high performance computing will therefore be able to
play a crucial role not only by contributing to scientific progress,
but also in its day-to-day use as a research and development tool in
mechanical and process engineering as well as biotechnology.

\begin{acknowledgement}
The present work was carried out under the auspices of the
Boltzmann-Zuse Society for Computational Molecular Engineering (BZS),
and the molecular simulations were conducted on the XC4000
supercomputer at the Steinbuch Centre for Computing, Karlsruhe.
The authors would like to thank Akshay Bedhotiya (Bombay) for his
assistance regarding the simulation of gas bubbles and Sergey Lishchuk
(Leicester), Martin Buchholz, Wolfgang Eckhardt,
and Ekaterina Elts (M\"unchen) and G\'abor Rutkai (Paderborn) as well
as Martin Bernreuther, Colin Glass, and Christoph Niethammer
(Stuttgart) for contributing to the development of
the \textit{ls1 mardyn} and \textit{ms2} molecular simulation
programs as well as Deutsche Forschungsgemeinschaft (DFG) for funding
the Collaborative Research Centre (SFB) 926. Furthermore, they would
like to acknow\-ledge fruitful discussions with Cemal Engin, Michael
Kopnarski, Birgit and Rolf Merz as well as Michael Schappals and
Rajat Srivastava (Kaiserslautern), George Jackson and Erich
M\"uller (London), Jonathan Walter (Ludwigshafen), Philippe Ungerer
and Marianna Yiannourakou (Paris) as well as Nichola McCann (Visp).
\end{acknowledgement}

% \bibliographystyle{spphys}
% \bibliography{mocos2013}

\end{document}